
\relax
\magnification=1200
\tolerance=1200

\baselineskip=18pt plus 2pt
\parindent=0pt
\parskip=.3cm

\def\IR{{\rm I\!R}}  
\def\IN{{\rm I\!N}}  

\centerline{\bf QUANTIZATION OF 2+1 GRAVITY ON THE TORUS}

\centerline{Adriana Criscuolo, Hernando Quevedo, and
Henri Waelbroeck}

\centerline{Instituto de Ciencias Nucleares}

\centerline{Universidad Nacional Aut\'onoma de M\'exico}

\centerline{A.P. 70--543, M\'exico D.F. 04510, MEXICO}
\vskip2cm
\centerline{\bf ABSTRACT}

We use the polygon representation of 2+1--dimensional gravity
to explicitly carry out the canonical quantization of a
universe with the topology of a torus. The mapping-class-invariant
wave function for a quantum ''big bounce'', is
reminiscent of the interference patterns of linear gratings.
We consider the ``problem of time'' of quantum gravity: for one
choice of internal time the universe recovers a semiclassical
interpretation after the bounce, with a wave packet centered
at a single geometry; for another choice of internal time,
the quantum solutions involve interference between macroscopically
distinct universes.

\vfill\eject
{\bf 1. Introduction}

As a consequence of Einstein's equations, the spacetime $M$ of
2+1--dimensional gravity in the absence of matter sources is flat.
If the universe is a Riemannian surface $\Sigma$ of genus $g$, there are
topological degrees of freedom related to the rotation of gyroscopes
which travel around non-contractible loops, and to the metric
variables which represent the sizes of these loops. One of the
interesting particularities of 2+1 gravity is that almost all
classical solutions have either initial or final singularities [1],
so the question arises whether the quantum generalizations of
these solutions can satisfactorily handle the singularities.

For a universe $\Sigma$, the spacetime manifold
$M$ is $\Sigma \times \IR$; let $T$ be its tangent bundle and
$T'$ a 3--dimensional vector bundle topologically equivalent to
$T$ and containing the structure group $SO(2,1)$. The equivalence
between $T$ and $T'$ implies the existence of at least one
isomorphism $e_i^{\ a}$ which in this case corresponds to a
dreibein, where $i$ and $a$ represent the tangent space and
Lorentz indices, respectively. In general, the dreibein
$e_i^{\ a}$ is assumed to be invertible in order for the metric
tensor in $M$ to be non degenerate. Nevertheless, there exist
regions in spacetime where $e_i^{\ a}$ is not everywhere
invertible. Those regions are of special interest in general
relativity as they correspond to ``classical singularities"
or regions of ``zero volume". In 2+1 quantum gravity Witten showed
that one must include the singular solutions if the theory is
to be renormalizable [2].

Among the various approaches to 2+1 quantum gravity, there are
essentially three different formulations: (i) The ``frozen time''
formulation [2, 3, 4] is based on the first--order form of the
Einstein--Hilbert action (Chern--Simons action); the
holonomy invariants of flat connections lead to a complete set of
Heisenberg observables;
(ii) The Arnowitt--Deser--Misner (ADM) formulation [5, 6] with
York's extrinsic time;
(iii) The polygon representation [7], which provides an explicit
representation of the phase space, which is particularly convenient
to consider dynamical issues. In this work, we will use this
representation and develop the quantization in the Schr\"odinger
picture.

We will consider only the simplest topological structure for $\Sigma$,
namely that of a torus. In section 2, we review the polygon representation
for a torus and present the generators of the mapping class group.
Section 3 is devoted to quantization. In Section 4, we examine in
particular the quantum analogue of ``big bounce'' solutions, where
the universe collapses to a singularity and then re-expands.

{\bf 2. Polygon representation}

A torus can be represented as a parallelogram with opposite edges
identified. Let us denote by $E(1)$ and $E(2)$ the three--vectors
which represent two of the edges in a Minkowski
embedding, and $M(1)$ and $M(2)$ the Lorentz
matrices used to identify the opposite sides $-M^{-1}(1)E(1)$ and
$-M^{-1}(2)E(2)$, respectively. The reduced phase space
can be parametrized by $E(\mu)$ and $M(\mu)$, $\mu=1,2$,
with the following non--vanishing Poisson brackets
($a,\ b, ... \in \{0,1,2 \},\  \epsilon^{012}= -1)$:
$$\{E^a(\mu), E^b(\mu)\} = \epsilon^{abc} E_c(\mu)\ , \qquad
\{E^a(\mu), M^b_{\ c}(\mu)\} = \epsilon^{abd} M_{dc}(\mu)\ .
\eqno(1)$$
The dynamics is generated by the
constraints
$$ J = E(1) + E(2) - M^{-1}(1)E(1) - M^{-1}(2)E(2) \approx 0 \ ,\eqno(2)$$
which imply that  the polygon closes, and the constraints
$$P^a = {1\over 2} \epsilon^{abc} W_{cb} \approx 0 \ , \eqno(3) $$
with $W = M(1) M^{-1}(2) M^{-1}(1) M(2)$, which represent the
cycle condition for the identification matrices.

There is an additional condition that must be taken into account.
Not all parallelograms generate different tori, but some of them
are equivalent under transformations of the mapping class group,
i.e., the group of isomorphisms that cannot be smoothly deformed
to the identity. For the torus, there are two different generators
of mapping class group transformations. The first one
$$E(1) \rightarrow E(1) + E(2) ,
\qquad E(2) \rightarrow M^{-1}(1) E(2) , $$
$$M(1) \rightarrow M(1) ,
\qquad M(2) \rightarrow M^{-1}(1) M(2) \eqno(4)$$
corresponds to a displacement along the base vector $E(2)$ such
that the new vector $E(1)$ coincides with the diagonal of the
initial parallelogram. For the second generator
$$E(1) \rightarrow E(1) ,
\qquad E(2) \rightarrow M(2)M^{-1}(1)E(1) + E(2),$$
$$M(1) \rightarrow M(1)M^{-1}(2) ,
\qquad M(2) \rightarrow M(2) \eqno(5)$$
the displacement is along the vector opposite to $E(1)$.

The Poisson brackets (1), the polygon closure relations
(2) and the cycle conditions (3), are all invariant under mapping
class transformations.

{\bf 3. Reduced phase space quantization}

{\bf A. Canonical variables}

The reduced phase space is defined by the variables $E^a(\mu)$ and
$M^a(\mu)$. To carry out the quantization programme, it is
convenient to introduce the canonical variables $X^a(\mu)$ and
their corresponding ``momenta" $P^a(\mu)$ by means of the
regular transformation
$$
X(\mu) = {1\over P^2(\mu)}\left[ P(\mu)\wedge J(\mu)
+2 { [E(\mu) . P(\mu)] P(\mu)\over {\hbox{tr}} M(\mu) -1 }
\right]\ ,
\eqno(6) $$
and
$$  P^a(\mu) = {1 \over 2} \epsilon^{abc} M_{cb}(\mu) \ , \eqno(7)$$
where $J(\mu)=[I-M^{-1}(\mu)] E(\mu)$ and the exterior product is
defined by $(A\wedge B)^a = \epsilon^{abc}A_c B_b $ (some of
the sign conventions differ from those of ref. [7]). The
Poisson brackets of these new variables are canonical
$\{P^a(\mu), X_b(\mu)\} = \delta^a_b $. The closure condition
may be written as
$$J(\mu) = \sum_\mu X(\mu)\wedge P(\mu) \approx 0\ , \eqno(8)$$ while the
translation constrains $P\approx 0$ are
given implicitly in terms of $P(\mu)$. The inverse transformation
can be calculated explicitly from Eqs.(5) and (6) [7]. To first order
in the momenta $P(\mu)$,
$$ M^a_{\ b}(\mu) = \delta^a_b + \epsilon^{ca}_{\ \ b} P_c (\mu) +
O(P^2)\ , \qquad    E(\mu) = X(\mu) - {1\over 2} X(\mu) \wedge
P(\mu) + O(P^2) \ . \eqno(9)$$

{\bf B. Mapping class transformations}

The canonical variables $X(\mu)$ and $P(\mu)$ defined above are
related to a particular choice of the base vectors $E(1)$ and $E(2)$
for the torus. Another choice of base vectors would lead to a different set
of canonical variables for the same spacetime. The different choices
are related by mapping class transformations.
A convenient representation of the mapping class group is achieved
by considering a parallelogram on $\IR^2$ with two basis vectors
$\hat e_1$ and $\hat e_2$. The mapping class images of this
parallelogram are
$\hat e'_1 = a_{11}\hat e_1 + a_{12} \hat e_2$
and $\hat e'_2 = a_{21}\hat e_1 + a_{22} \hat e_2$,
where $a_{11}, \cdots \in \IN$ and $a_{11}  a_{22} -  a_{12}  a_{21} = 1$.
The corresponding tranformation for the polygon vectors embedded
in Minkowski space is:
$$\eqalignno{
E'(1) = & \ \left[1 + M(2) + \cdots + M^{a_{11}-1}(2) \right] E(1) \cr
& \  + \left[ 1+ M^{-1}(1) + \cdots + M^{-a_{12}+1}(1)
\right] M^{a_{11}-1}(2) E(2) \ , \ & (10) \cr
E'(2) = & \ \left[1 + M^{-1}(1) + \cdots + M^{-a_{22}+1}(1) \right] E(2) \cr
& \  + \left[ 1+ M(2) + \cdots + M^{a_{21}-1}(2)
\right] M^{-a_{22}}(1) M(2) E(1) \ , \ & (11) \cr
M'(1) = & \ M^{-a_{21}}(2)M^{a_{22}}(1)\ , \ & \  (12) \cr
M'(2) = & \ M^{-a_{12}}(1)M^{a_{11}}(2)\ . \ & \  (13) }$$

With these relations, one can compute the action of the mapping
class group on the canonical variables. To first order in $P$,
$$\eqalignno{
X'(1) = & \ a_{11} X(1) + a_{12} X(2) - {a_{11}\over 2}
[(a_{22}-1) P(1) - (a_{11}+a_{21} - 1) P(2)] \wedge X(1) \cr
& \ - {a_{12}\over 2}
[(a_{12}+a_{22}-1) P(1) - (2a_{11}+a_{21} - 1) P(2)] \wedge X(2)
+ O(P^2) \ , & (14)  \cr
X'(2) = &\ a_{21} X(1) + a_{22} X(2)
-  {a_{21}\over 2}
[(2 a_{22}- a_{12} -1) P(1) + (a_{11}-a_{21} - 1) P(2)] \wedge X(1) \cr
& \ -  {a_{22}\over 2}
[(a_{22}-a_{12}-1) P(1) + (a_{11} - 1) P(2)] \wedge X(2)
+ O(P^2) \ . & (15) }$$
These expressions will be needed to calculate the mapping class invariant
wave function, below.

{\bf C. The internal time}

Our goal is the calculation of wave functions for the torus
``big bounce''. First we have to choose one of the variables as
``internal time'' and calculate the corresponding Hamiltonian.
Since we are not considering punctures on the torus, the Lorentz
matrices $M(\mu)$ are pure boosts. Without loss of generality
we can choose a frame in which the momentum $P(1)$ is parallel to
a spatial axis, say $x$. Then
$$P^t(1) = P^y(1) = 0 \eqno(16)$$
and
$$
M^a_{\ b}(1) = \left(\matrix{ \cosh\beta & 0 & \sinh\beta \cr
                          0      & 1 & 0 \cr
                      \sinh\beta & 0 & \cosh\beta \cr}\right)\ ,
\eqno(16)$$
where $\beta$ is the boost parameter. These conditions fix the
``gauge symmetries'' generated by the constraints $J^t = J^y =0$, which
must be solved for the variables conjugate to the gauge conditions:
$$X^t(1) = {X^x(2) P^t(2) - X^t(2)P^x(2)\over P^x(1)}\ ,
\qquad
X^y(1) = {X^x(2) P^y(2) - X^y(2)P^x(2)\over P^x(1)} \ .
\eqno(17)$$
The further constraints $P\approx 0$ become
$$P^y(2)\approx 0\ , \qquad  P^t(2) \approx 0\ .\eqno(18)$$
It follows from the constraint $J^x=0$ that $X^t(2)=0$. With the
remaining components of the canonical variable $X(\mu)$,
one can construct the internal time which must be a variable
that does not commute with the constraints (18). The choice
$$ t= \cos\alpha\, X^y(2) - \sin\alpha\, X^x(2) \eqno(19)$$
satisfies this criterion, where $\alpha$ is a nonvanishing
parameter. Defining new variables
$x_\mu$  and $p_\nu$ ($\mu, \nu \in \{ 1,2 \} $) with canonical
Poisson brackets $\{x_\mu, p_\nu\} = \delta_{\mu\nu}$ in the following way
$$x_1 = X^x(1)\ , \quad p_1 = P_x(1)\ , \quad
x_2 = \cos\alpha\, X^x(2) + \sin\alpha\, X^y(2)\ , \quad
p_2 = P_x(2)/\cos\alpha \ , \eqno(20)$$
the Hamiltonian corresponding to the internal time (19) can be
written as
$$ H = -\tan\alpha\, p_2\ . \eqno(21)$$

{\bf D. The wave function}

The calculation of wave functions implies two different aspects:
(1) the dynamical problem, i.e. determination of the Hamiltonian,
and (2) the mapping class problem, i.e. finding expressions which
are invariant with respect to transformations of the mapping class
group. The first problem has been solved for the torus in the last
subsection. Here we will attack the second problem.

Consider a wave packet described by the function $\tilde a (\vec x _T)$
at some initial time $T$. The propagation of
this packet from $T$ to an arbitrary time $t$ can be described
by a wave function involving a path integral of the form
$$
\tilde\psi (\vec x_t, t) = \int d\vec x _T
\tilde a(\vec x _T)\int {\cal D} \vec x(\tau) {\cal D}
\vec p(\tau)
 \, e^{ i \int^t_T (\vec p\, \dot{\vec x} - H)
d\tau } \delta(G)\delta(K) |Det\{G, K\}| \ , \eqno(22)$$
where $G\approx 0$ are the gauge constraints, $K\approx 0$
represent the gauge conditions, and $|Det\{G, K\}|$ is the
determinant of the corresponding Jacobian. Splitting up
the interval $t-T$ into $N$ infinitesimal time intervals,
the expression $\int {\cal D}\vec x {\cal D}\vec p$ includes
$N-1$ integrations over $d<x_i>$, where
$<x_i>$ is the main value of $x(\tau_i)$ in each
infinitesimal interval, and $N$ integrations $d\vec p_T \ ...
\ d\vec p_t$. Since $H$ is a function of $p$ only, the
integration over each $<x_i>$ yields a term of the form
$\delta(\vec p_{i+1} - \vec p_i)$. The further
integration over $d\vec p_T \ ... \ d\vec p_t$
leads to
$$
\tilde\psi (\vec x_t, t) = C \int d\vec x _T
\tilde a(\vec x _T)\int d \vec p_T
 \, e^{ i [\vec p_T (\vec x_t - \vec x_T)
- H\,  (t-T)] }
 \delta(G)\delta(K) |Det\{G, K\}| \ , \eqno(23)$$
where $C$ is a constant, and the Hamiltonian $H$ is evaluated
at $\vec p_T$.

The expression (23), however, is not invariant with respect to
mapping class transformations. To reach this invariance, we
apply the method of images, i.e. we consider the sum over
all mapping class images of $x$ (including $x_0 = t$). Then
$$\psi_{inv} (\vec x_t, t) \sim \sum_\gamma \hat O_\gamma
\tilde\psi (\vec x_t, t)\ , \eqno(24)$$
where the sum is over all transformations ($\gamma)$ of the
mapping class group, and $\hat O_\gamma $ is an operator
which represents the action of each transformation on the
non-invariant wave function. To determine the action of
$\hat O_\gamma $ on $\tilde\psi$, it is convenient to
transform the scalar products entering Eq.(23) into
covariant expressions. This can be done by multiplying
the integrand in (23) by $\delta(p_0 - H)\, \delta (x_0 -t)$
and changing  $ d\vec x _T  d \vec p_T$ by
$ d x_T  d p_T$. Clearly, this change does not alter the result
of the integration, but allows us to write the argument of the
exponential in Eq.(23) in a covariant way: $ p_T ( x_t - x_T)$.
On the resulting wave function we apply the operator
$\hat O_\gamma$ that modifies $x$ into $x'_\gamma$, and then
integrate over $x'^0_\gamma$, with the factor $\delta(x'^0_\gamma - t)$;
thus, the parameter $t$ is invariant with respect to mapping
class transformations, and the invariant wave function (24) can be written as
$$\psi_{inv} (\vec x_t, t)\sim \sum_\gamma
\int d\vec x _T d \vec p_T
\tilde a(\vec x _T)
 \, e^{ i [\vec p_T (\vec x'_t - \vec x_T)
- H\,  (t-T)] }
 \delta(G_\gamma)\delta(K_\gamma) |Det\{G_\gamma, K_\gamma\}| \ , \eqno(25)$$
where the index $\gamma$ was omitted in $x'_\gamma$ for the
sake of clarity.

To proceed with the integration of the wave function (25), we specify
the initial wave packet as a Gaussian distribution, at $T = 0$:
$$\tilde a \sim e^{ -{(x^1_T-a)^2 + (x^2_T-b)^2\over 2\sigma^2} }
e^{i \vec k \cdot \vec x}\ , $$
where $a$, $b$, $\vec k$ and $\sigma$ are constants. The explicit values
of the mapping class images $\vec x '$ can be obtained by
using the defining equations (20) and the relations (14) and
(15), which represent the action of mapping class transformations
on the original canonical variables $X(\mu)$. The resulting
wave function is given by
$$\eqalignno{
& \psi_{inv}(\vec x_t, t) \sim  \sum_\gamma
\int dp_1 dp_2 e^{ -{\sigma^2\over 2}((p_1 - k_1)^2+(p_2 - k_2)^2) } \cr
& \ \times
 e^{ i [ (p_1-k_1)(x_1 -a + \Delta x_1)
+ (p_2 - k_2)(x_2 -b+ \Delta x_2 + \tan\alpha\, t)] }
   \delta(G_\gamma)\delta(K_\gamma) |Det\{G_\gamma, K_\gamma\}|
\ , & (26) \cr }
$$
where
$$\eqalign{
\Delta x_1 = &\  (a_{11} - 1) x_1 + a_{12} x_2 \cos\alpha
- a_{12} t \sin\alpha + O(p^2) \ , \cr
\Delta x_2 = &\  (a_{22} - 1) x_2 + a_{21} x_1 \cos\alpha
- a_{21} \sin\alpha \cos\alpha\, (x_2 \sin\alpha\, + t \cos\alpha\, )
{p_2\over p_1} + O(p^2)\ .\cr}
$$
The sum in Eq.(26) is over all integer values of $a_{ij}$ such
that $a_{11}  a_{22} -  a_{12}  a_{21} = 1$. This wave
function (26) describes the propagation of an initial
packet $\tilde a (\vec x_T)$; we will consider in
particular the case where this packet is centered about a collapsing
universe, and examine the quantum ``big bounce''.

{\bf 4. Conclusions: The ``Big Bounce'', and the Problem of Time}

Let us consider the initial packet centered at $\{ a, b \}$, with
initial momentum distributed about $\vec k = \{ k_1, k_2 \}$. The classical
evolution is given by the dynamical equations ${\buildrel \cdot \over x_1}
 = 0$, ${\buildrel \cdot \over x_2} = - \tan{\alpha}$, so
the classical trajectory goes through a singular universe $\{a , 0 \}$
at time $t = {b \over \tan\alpha}$ and then unfolds into the universe
$\{ a, -b \}$ at $t = {2b \over \tan\alpha}$. There can be other classical
trajectories from $\{ a, b \}$ to $\{ a, -b \}$: indeed, the final value
of the variables, $\{ a, -b \}$, might well correspond to a different
set of basis loops; correcting for the mapping class transformation
between the initial and final states, we would then look at the
classical evolution from $\{ a, b \}$ to
$$\{ a + \Delta x_1, \ - b + \Delta x_2 \}\ , \eqno(27)$$
where $\Delta x_i$ are evaluated at $\vec x = \{ a, -b\}$.
Since ${\buildrel \cdot \over x_1} = 0$, this can be a classical
trajectory only if $a_{12} = 0$ and $a_{11} = 1$, which implies that
$a_{22} = 1$. There is then one classical trajectory for each $a_{21}$, which
reaches the ``detector'' (of universes...) at  $\{ a, -b \}$ in a
time given implicitly by $t \tan\alpha = 2b - \Delta x_2$. Conversely,
at time $t = {2b \over \tan\alpha}$ this classical solution reaches
$$ x_2 = -b - \Delta x_2 = -b - a_{21} ( a \cos\alpha
- {\sin{2\alpha} \over 2} ( - b \sin\alpha\ + t \cos\alpha\ )
{k_2\over k_1} ). \eqno(28)$$

In the quantum theory one has a sum of amplitudes, and the question arises
of whether the amplitudes interfere. Since $\Delta x_2 \sim a_{21} a$ and
$a$ is a {\it macroscopic} constant of the motion, one does not expect
interference of images with different values of $a_{21}$; so only one of
the classical trajectories described above will contribute. On the other
hand, with $\Delta x_1 \sim a_{12} x_2$, one finds that the images for
various values of $a_{12}$ are distributed as in a linear grating.
Therefore, one expects interference to occur when the variance of
the wave packet is at least of the same order of magnitude as the
spacing $\eta$ between slits of the ``grating'',
$$\eta = x_2 \cos\alpha\ . \eqno(29)$$
This indicates that quantum interference is surely going to be
significant near the singularity, where $x_2 \simeq 0$.

One might worry about the summability of the expression for the invariant
wave function in the limit $x_2 \to 0$, as the number of terms in the sum
grows without bound -- this is related to the delicate issue of how the
quantum theory handles the singularity. However, the
wave function is nothing but the unitary evolution of the initial
packet with a smooth Hamiltonian, so by construction
it is normed at all times; the challenge would be to find the appropriate
regularization of the sum near $x_2 = 0$, or to analytically continue an
expression which is computable for all $x_2 \neq 0$.

Another question, which we will consider in greater detail, is whether
there can be interference between macroscopically distinct universes after
the ``big bounce''.

Since we have chosen an internal time which leads
to a Hamiltonian linear in $p_2$, there is no dispersion and therefore
the variance is equal to the initial variance $\sigma^2$ at all times.
Assuming that this initial variance is microscopic, one finds
interference effects only near the singularity.

What if one had chosen a different internal time? Consider for instance
$$t' = {{\cos\alpha \ X^y(2) - \sin\alpha \ X^x(2)} \over {-2 \sin\alpha
\ P_x(2)}} \ . \eqno(30)$$
The corresponding Hamiltonian is
$$H = \tan^2 \alpha \ p_2^{\ 2} \ , \eqno(31)$$
which leads to the dispersion $\sigma^2(t) = \sigma^2 + \tan^2 \alpha\ t'$.
Thus, after a sufficiently large amount of time (such as the time
to go through the singularity and re-expand to a large universe), the
variance becomes large and one has interference between macroscopically
distinct universes.

This is a particularly striking manifestation of one of the ``problems
of time'' of quantum gravity, known as the multiple choice problem:
different choices of internal time lead to different physical predictions.
In this case, the interference lines with the second
choice of time will lead to sizeable amplitudes for non-classical
histories. In contrast, with the first choice of internal time one
finds a significant amplitude only near the predictions
of the classical theory.

{\bf References}

\item{[1]} G. Mess, Lorentz spacetime of constant curvature,
Institut des Hautes Etudes Scientifiques Preprint IHES/M/90/28 (1990)

\item{[2]} Ed Witten, Nucl. Phys. B311 (1988) 46

\item{[3]} S. Carlip, Phys. Rev. D45 (1992), 3584

S. Carlip, Class. Quantum Grav.8 (1991), 5

S. Carlip, Phys. Rev. D42 (1990), 2647

S. Carlip, Nucl. Phys. B324 (1989), 106

\item{[4]} J. E. Nelson and T. Regge, Commun. Math. Phys. 141
(1991), 211

J. E. Nelson and T. Regge, Phys. Lett. B272 (1991), 213

J. E. Nelson, T. Regge and F. Zertuche, Nucl. Phys. B339 (1990), 516

J. E. Nelson, T. Regge, Nucl. Phys. B328 (1989), 190

\item{[5]} V. Moncrief, J. Math. Phys. 31 (1990), 2978

V. Moncrief, J. Math. Phys. 30 (1989), 2297

\item{[6]} A. Hosoya and K. Nakao, Class. Quant. Grav. 7 (1990), 63

\item{[7]} H. Waelbroeck and F. Zertuche, Phys. Rev. D50 (1995), 4966

H. Waelbroeck, Phys. Rev. D50 (1995), 4982

H. Waelbroeck, Nucl. Phys. B364 (1991), 475

\bye